# An Analytic Boris Pusher for Plasma Simulation

Viktor K. Decyk, Warren B. Mori, and Fei Li

Department of Physics and Astronomy and Department of Electrical Engineering
University of California, Los Angeles, Los Angeles, CA 90095, USA.

Abstract

This paper discusses how to improve the Boris pusher used to advance relativistic charged particles in fixed electromagnetic fields. We first derive a simpler solution to a flaw previously discovered by others. We then derive a new analytic Boris pusher that is a minor modification to the original split-time scheme, except for the calculation of $\gamma$. This analytic pusher assumes that the change of $\gamma$ during a particle advance is small. It is less accurate than the pusher derived by Fei Li et. al., but is nearly twice as fast. We will discuss when it is advantageous and when it is not.

## 1.0 Introduction

One of the most important and long-lived algorithms in plasma physics is the Boris algorithm [1] for solving the relativistic equations of motion for charged particles in electromagnetic fields. It has been in use in Particle-in-Cell (PIC) simulations [2,3] for more than 50 years. It solves the equations:

$$\frac{d\mathbf{u}}{dt} = \frac{q}{m}\left[\mathbf{E} + \mathbf{v} \times \frac{\mathbf{B}}{c}\right] \text{ and } \frac{d\mathbf{x}}{dt} = \mathbf{v} \tag{1.}$$

where we define:

$$\mathbf{u} = \gamma\mathbf{v} \text{ and } \gamma = \sqrt{1 + \mathbf{u} \cdot \mathbf{u}/c^2} \tag{2.}$$

We use boldface here to indicate vectors. The Boris algorithm assumes that the relativistic factor $\gamma$ and the electric and magnetic fields $\mathbf{E}$ and $\mathbf{B}$ do not vary during a time step $\Delta t$. The algorithm uses a time-splitting scheme with 4 parts.

First, the particle is accelerated half a time step with the electric field **E** only:

$$\mathbf{u1} = \mathbf{u} + \frac{q\Delta t}{2m}\mathbf{E} \tag{3.}$$

Second, the $\gamma$ factor is calculated:

$$\gamma_B = \sqrt{1 + \mathbf{u1} \cdot \mathbf{u1}/c^2} \tag{4.}$$

Third, the particle is rotated with the magnetic field **B** only using the equation:

$$\mathbf{u2} = \mathbf{u1} + \frac{q\Delta t}{2m\gamma_B c}(\mathbf{u1} + \mathbf{u2}) \times \mathbf{B} \tag{5.}$$

This implicit equation can be solved for **u2** either by inverting a 3x3 matrix or by using an optimized scheme given in [1]. The result can be written in the following vector form:

$$\mathbf{u2} = \left\{\mathbf{u1}\left[\mathbf{1} - \left(\frac{\Omega\Delta t}{2\gamma_B}\right)^{\mathbf{2}}\right] + \frac{\mathbf{u1} \times \mathbf{\Omega}\Delta t}{\gamma_B} + \frac{1}{2}\left(\frac{\Delta t}{\gamma_B}\right)^2 (\mathbf{u1} \cdot \mathbf{\Omega})\mathbf{\Omega}\right\} \Big/ \left[\mathbf{1} + \left(\frac{\Omega\Delta t}{2\gamma_B}\right)^{\mathbf{2}}\right] \tag{6.}$$

where $\mathbf{\Omega} = \frac{q\mathbf{B}}{mc}$. Finally, the particle is accelerated another half a time step with the electric field **E** only:

$$\mathbf{u} = \mathbf{u2} + \frac{q\Delta t}{2m}\mathbf{E} \tag{7.}$$

It is noted in [1] that the rotation can be made more exact by replacing

$$\frac{\Delta t}{2} \rightarrow \frac{\gamma_a}{\Omega}\tan\left(\frac{\Omega\Delta t}{2\gamma_B}\right) \tag{8.}$$

Although this replacement improves the accuracy of the rotation, it does not improve the overall solution very much and is rarely used. More than a decade ago, a flaw was discovered [4] in this algorithm when moving relativistic particles where $\mathbf{E} + \mathbf{v} \times \mathbf{B}/c \approx 0$. A number of solutions were proposed [4,5] to fix this.

In more recent years, the development of high-intensity lasers has stimulated the need to simulate plasmas with large electromagnetic fields. It was observed that the Boris algorithm with such fields requires small time steps to give accurate results [6-7]. As a result, a more accurate analytic pusher was developed by Fei Li et al. [8] which allows larger time steps. This pusher assumes that $\mathbf{E}$ and $\mathbf{B}$ are constant during a time step, as in the Boris algorithm, but $\gamma$ is now allowed to vary. This pusher is more complex and slower than the Boris algorithm, but the overall performance can be better due to the larger time steps possible for fields which are strong enough. The purpose of this paper is to develop an analytic pusher where $\gamma$ is constant as in the Boris algorithm, but more accurately calculated. It is intended to occupy a middle ground between the more accurate but slower scheme of Fei Li and the less accurate but faster Boris algorithm for fields which are less strong. We will derive the various solutions and discuss when they are advantageous. Unlike [8], the paper will make use of the more traditional 3-vector plus time representation.

## 2.0 Correction to the Boris Pusher

We will begin by examining the accuracy of the Boris gamma factor $\gamma_B$. From the definition of $\gamma$, equation (2), and the equations of motion, equation (1), one can show that:

$$\frac{d\gamma}{dt} = \frac{1}{2\gamma c^2}\frac{d}{dt}(\mathbf{u}\cdot\mathbf{u}) = \frac{q}{\gamma m c^2}\mathbf{E}\cdot\mathbf{u} \qquad (9.)$$

The classic Boris algorithm does not make use of this equation directly. Multiplying by $2\gamma$, one can then show that:

$$\frac{d\gamma^2}{dt} = \frac{2q}{mc^2}\mathbf{E}\cdot\mathbf{u} \;\Rightarrow\; \frac{d^2}{dt^2}(\gamma^2) = \frac{2q}{mc^2}\,\mathbf{E}\cdot\frac{d\mathbf{u}}{dt} \qquad (10.)$$

From these equations one can express the value of $\gamma^2(t)$ as a Taylor series:

$$\gamma^2\left(\frac{\Delta t}{2}\right) = \gamma^2(0) + \frac{q}{mc^2}\mathbf{E}\cdot\mathbf{u}(\mathbf{0})\Delta t + \frac{q}{mc^2}\,\mathbf{E}\cdot\frac{d\mathbf{u}(\mathbf{0})}{dt}\frac{\Delta t^{\mathbf{2}}}{4} \qquad (11.)$$

Compare this with the Boris algorithm, equation (4):

$$\gamma_B^2 = 1 + \left(\mathbf{u}(\mathbf{0}) + \frac{q\Delta t}{2m}\mathbf{E}\right)\cdot\left(\mathbf{u}(\mathbf{0}) + \frac{q\Delta t}{2m}\mathbf{E}\right)/c^{\mathbf{2}} \qquad (12.)$$

$$= \gamma^2(0) + \frac{q\Delta t}{mc^2}\mathbf{E}\cdot\mathbf{u}(\mathbf{0}) + \frac{q\Delta t^2}{4mc^2}\mathbf{E}\cdot\frac{q}{m}\mathbf{E} \qquad (12.)$$

One can see that the Boris $\gamma_B^2$ does not agree with $\gamma^2(\frac{\Delta t}{2})$ because $\frac{d\mathbf{u}(\mathbf{0})}{dt}$ includes the magnetic force and $\gamma_B$ does not. If the magnetic force is small compared with the electric force, the difference is small. A correction can easily be added by retrieving the magnetic force term in $\frac{d\mathbf{u}(\mathbf{0})}{dt}$. If one defines:

$$\delta\gamma_B^2 = \frac{q\Delta t^2}{4mc^2}\mathbf{E}\cdot\mathbf{u}(\mathbf{0})\times\frac{\boldsymbol{\Omega}}{\gamma(0)} = -\frac{q\Delta t^2}{4mc^2}\mathbf{u}(\mathbf{0})\cdot\mathbf{E}\times\frac{\boldsymbol{\Omega}}{\gamma(0)} \qquad (13.)$$

then $\gamma_C$ below gives a second order accurate expression for the Boris $\gamma$ factor:

$$\gamma_C = \sqrt{\gamma_B^2 + \delta\gamma_B^2} \qquad (14.)$$

This correction requires an additional square root to calculate the initial $\gamma(0)$, and the overall increase in computational time in a 2D PIC code was about a factor of 1.3. If $\mathbf{E}\cdot\boldsymbol{\Omega} = 0$ and $\mathbf{u}(0) = \gamma(0)\frac{q}{m\Omega^2}\mathbf{E}\times\boldsymbol{\Omega}$, then one can show that $\gamma_C = \gamma(0)$ and $\mathbf{u}(t) = \mathbf{u}(0)$, which also fixes the flaw previously reported in [4,5].

## 3.0 Analytic Boris Pusher

### 3.1 Solution to Equations of Motion

We will now proceed to solve the Boris problem analytically for fixed $\mathbf{E}$, $\mathbf{B}$, and $\gamma$. First, one separates the equations of motion, equation (1), into two parts, parallel and perpendicular to $\mathbf{B}$. If one defines the unit vector in the direction along the magnetic field as $\widehat{\Omega} = \frac{\mathbf{\Omega}}{\Omega}$, then one can write:

$$\frac{d\mathbf{u}_{\parallel}}{dt} = \frac{q}{m}\mathbf{E}_{\parallel} \tag{15.}$$

where $\mathbf{E}_{\parallel} = (\mathbf{E} \cdot \widehat{\Omega})\widehat{\Omega}$ and $\mathbf{u}_{\parallel} = (\mathbf{u} \cdot \widehat{\Omega})\widehat{\Omega}$ and

$$\frac{d\mathbf{u}_{\perp}}{dt} = \frac{q}{m}\mathbf{E}_{\perp} + \mathbf{u} \times \frac{\mathbf{\Omega}}{\gamma_a} \tag{16.}$$

where $\mathbf{E}_{\perp} = \mathbf{E} - \mathbf{E}_{\parallel}$ and $\mathbf{u}_{\perp} = \mathbf{u} - \mathbf{u}_{\parallel}$. The symbol $\gamma_a$ represents some average value of $\gamma$ to be determined. The solutions are:

$$\mathbf{u}_{\parallel}(t) = \mathbf{u}_{\parallel}(0) + \frac{q}{m}\mathbf{E}_{\parallel}t \tag{17.}$$

$$\mathbf{u}_{\perp}(t) = [\mathbf{u}_{\perp}(0) - \gamma_a\mathbf{v}_E]\cos\left(\frac{\Omega t}{\gamma_a}\right) + \frac{1}{\Omega}\left[\mathbf{u}_{\perp}(0) \times \mathbf{\Omega} + \frac{q\gamma_a}{m}\mathbf{E}_{\perp}\right]\sin\left(\frac{\Omega t}{\gamma_a}\right) + \gamma_a\mathbf{v}_E \tag{18.}$$

where $\mathbf{v}_E = \frac{q}{m\Omega^2}\mathbf{E}_{\perp} \times \mathbf{\Omega}$. The correctness of this solution can be verified by substituting these expressions into the equations of motion (15-16). It can also be derived directly from the equations of motion using Laplace transform techniques.

If one defines the unit vector $\widehat{E}_{\perp} = \frac{\mathbf{E}_{\perp}}{E_{\perp}}$, then one can write down the solutions (17-18) as 3 scalar equations by taking the dot product of $\mathbf{u}$ and $\mathbf{E}$ with $\widehat{\Omega}$, $\widehat{E}_{\perp}$, and $\widehat{E}_{\perp} \times \widehat{\Omega}$, respectively:

$$u_{\parallel}(t) = u_{\parallel}(0) + \frac{q}{m}E_{\parallel}t \tag{19.}$$

$$u_L(t) = u_L(0)\cos\left(\frac{\Omega t}{\gamma_a}\right) + \frac{1}{\Omega}\left[\frac{q\gamma_a}{m}E_{\perp} - u_D(0)\Omega\right]\sin\left(\frac{\Omega t}{\gamma_a}\right) \tag{20.}$$

$$u_D(t) = [u_D(0) - \gamma_a v_E]\cos\left(\frac{\Omega t}{\gamma_a}\right) + u_L(0)\sin\left(\frac{\Omega t}{\gamma_a}\right) + \gamma_a v_E \tag{21.}$$

where $u_L = \mathbf{u} \cdot \widehat{E}_{\perp}$ and $u_D = \mathbf{u} \cdot \widehat{E}_{\perp} \times \widehat{\Omega}$. These equations represent gyromotion whose center is accelerating in the $\widehat{\Omega}$ direction, while moving in the $\widehat{E}_{\perp} \times \widehat{\Omega}$ direction with velocity $v_E$, as expected.

3.2 Expression for $\gamma_a$

In order to calculate a suitable average value for $\gamma_a$, one needs to make use of proper time $\tau$ as defined by:

$$\frac{d\tau}{dt} = \frac{1}{\gamma} \tag{22.}$$

which leads to the relation between time and proper time intervals:

$$\Delta t = \int_0^{\Delta\tau} \gamma(\tau')d\tau' \tag{23.}$$

One can now define the average value of $\gamma$ as follows:

$$\gamma_a = \frac{1}{\Delta\tau}\int_0^{\Delta\tau} \gamma(\tau')d\tau' \Rightarrow \gamma_a = \frac{\Delta t}{\Delta\tau} \tag{24.}$$

To calculate the actual value of $\gamma_a$, one can make use of the equations (22) and (9) to obtain:

$$\frac{d\gamma}{d\tau} = \frac{\gamma d\gamma}{dt} = \frac{1}{c^2}\,\tilde{\mathbf{E}} \cdot \mathbf{u} \tag{25.}$$

where $\tilde{\mathbf{E}} = \frac{q}{m}\mathbf{E}$. Making use of equations (19-21) and substituting $t = \gamma_a\tau$, one can show that:

$$\frac{d\gamma}{d\tau} = \frac{\tilde{E}_\parallel}{c^2}\{u_\parallel(0) + \gamma_a\tilde{E}_\parallel\tau\} + \frac{\tilde{E}_\perp}{c^2}\left\{u_L(0)\cos(\Omega\tau) + \frac{1}{\Omega}\frac{du_L(0)}{d\tau}\sin(\Omega\tau)\right\} \tag{26.}$$

where $\frac{du_L(0)}{d\tau} = \gamma_a\tilde{E}_\perp - u_D(0)\Omega$. Integrating this with respect to $\tau$, one can show that:

$$\gamma(\tau) = \gamma(0) + \frac{\tilde{E}_\parallel}{c^2}\left\{u_\parallel(0)\tau + \frac{\gamma_a}{2}\tilde{E}_\parallel\tau^2\right\}$$
$$+\frac{\tilde{E}_\perp}{\Omega c^2}\left\{u_L(0)\sin(\Omega\tau) - \frac{1}{\Omega}\frac{du_L(0)}{d\tau}[\cos(\Omega\tau) - 1]\right\} \tag{27.}$$

Finally, integrating the expression for $\gamma(\tau)$ over the interval $\Delta\tau$, gives the result:

$$\Delta t(\Delta\tau) = \gamma(0)\Delta\tau + \frac{\tilde{E}_\parallel}{2c^2}\left\{u_\parallel(0)(\Delta\tau)^2 + \frac{\gamma_a}{3}\tilde{E}_\parallel(\Delta\tau)^3\right\}$$
$$-\frac{\tilde{E}_\perp}{\Omega^2 c^2}\left\{u_L(0)[\cos(\Omega\Delta\tau) - 1] + \frac{1}{\Omega}\frac{du_L(0)}{d\tau}[\sin(\Omega\Delta\tau) - \Omega\Delta\tau]\right\} \tag{28.}$$

This is a formal expression because one does not know the value of $\Delta\tau$ for which $\Delta t(\Delta\tau) = \Delta t$. One needs to solve the transcendental equation $\Delta t(\Delta\tau) - \Delta t = 0$ to determine the correct value of $\Delta\tau$ for a given $\Delta t$. It is analogous to equation 2.13 in [8], but this equation is simpler since it does not have hyperbolic terms. If one uses a Newton method, one can start with initial guesses:

$$\Delta\tau_1 = \frac{\Delta t}{\gamma(0)} \text{ and } \gamma_{a,1} = \gamma(0)$$

and iterate to the desired accuracy:

$$\Delta\tau_{n+1} = \Delta\tau_n - \frac{\Delta t(\Delta\tau_n) - \Delta t}{\gamma(\Delta\tau_n)} \text{ and } \gamma_{a,n+1} = \frac{\Delta t}{\Delta\tau_{n+1}} \tag{29.}$$

Note that $\gamma_a$ needs to be updated when $\Delta\tau$ is updated during the iteration. The iteration should take about half as much CPU time as the one in [8].

Since this analytic pusher assumes $\gamma$ varies little from one time step to another, high accuracy in the calculation of $\gamma_a$ may not be necessary. If the time step is small enough, one can approximate the expression for $\gamma(\tau)$ with a Taylor series:

$$\gamma(\tau) = \gamma(0) + \frac{d\gamma(0)}{d\tau}\tau + \frac{d^2\gamma(0)}{d\tau^2}\frac{\tau^2}{2}\cdots \quad (30.)$$

where $\frac{d\gamma(0)}{d\tau} = \frac{1}{c^2}\,\tilde{\mathbf{E}}\cdot\mathbf{u(0)}$ and $\frac{d^2\gamma(0)}{d\tau^2} = \frac{1}{c^2}\,\tilde{\mathbf{E}}\cdot\frac{d\mathbf{u(0)}}{d\tau}$ and then integrate in $\tau$ to obtain:

$$\gamma_a = \frac{\Delta t(\Delta\tau)}{\Delta\tau} = \gamma(0) + \frac{d\gamma(0)}{d\tau}\frac{\Delta\tau}{2} + \frac{d^2\gamma(0)}{d\tau^2}\frac{(\Delta\tau)^2}{6} + \cdots \quad (31.)$$

An improved value for $\gamma_a$ can be obtained by calculating $\Delta\tau = \frac{\Delta t}{\gamma_a}$, and substituting into equation (31) again. As discussed in section 4 below, the second order Taylor approximation appears adequate when $\frac{\Omega\Delta t}{\gamma_a} < 0.2$. Higher order derivatives can be calculated by repeatedly differentiating equation (26) and evaluating at $\tau = 0$. The third and fourth order coefficients are: $\frac{d^3\gamma(0)}{d\tau^2} = -\frac{\Omega^2}{c^2}\,\tilde{\mathbf{E}}_\perp\cdot\mathbf{u(0)}$ and $\frac{d^4\gamma(0)}{d\tau^2} = -\frac{\Omega^2}{c^2}\,\tilde{\mathbf{E}}_\perp\cdot\frac{d\mathbf{u(0)}}{d\tau}$. The fourth order Taylor approximation appears adequate when $\frac{\Omega\Delta t}{\gamma_a} < 0.8$.

### 3.3 Split-Time Formulation

One can also express the analytic Boris pusher as a split-time scheme as follows: First, calculate $\gamma_a$ as described in subsection 3.2 above, a significant change from the classic Boris algorithm. Second, accelerate the particle a half time step as follows:

$$\mathbf{u1'} = \mathbf{u(0)} + \frac{q}{m}\mathbf{E}_\parallel\frac{\Delta t}{2} + \frac{q\gamma_a}{m\Omega}\mathbf{E}_\perp\tan\left(\frac{\Omega\Delta t}{2\gamma_a}\right) \quad (32.)$$

Third, rotate the particle with the magnetic field **B** using the equation:

$$\mathbf{u2'} = \mathbf{u1'}\cos\left(\frac{\Omega\Delta t}{\gamma_a}\right) + \frac{\mathbf{u1'}\times\mathbf{\Omega}}{\Omega}\sin\left(\frac{\Omega\Delta t}{\gamma_a}\right) + \left[1-\cos\left(\frac{\Omega\Delta t}{\gamma_a}\right)\right]\frac{(\mathbf{u1'}\cdot\mathbf{\Omega})\mathbf{\Omega}}{\Omega^2} \quad (33.)$$

Finally, the particle is accelerated another half time step:

$$\mathbf{u} = \mathbf{u2'} + \frac{q}{m}\mathbf{E}_\parallel\frac{\Delta t}{2} + \frac{q\gamma_a}{m\Omega}\mathbf{E}_\perp\tan\left(\frac{\Omega\Delta t}{2\gamma_a}\right) \quad (34.)$$

Note that the half acceleration perpendicular to **B**, equation(32), is treated differently than in the classic Boris algorithm, equation (3). This requires that the **E** field must be decomposed into components parallel and perpendicular to the magnetic field. The rotation is the same as the Boris rotation, but requires the more exact version:

$$\frac{\Delta t}{2} \to \frac{\gamma_a}{\Omega}\tan\left(\frac{\Omega\Delta t}{2\gamma_a}\right) \quad (35.)$$

Carrying out the operations in equations (32-34), one can recover equations (17-18).

3.4 Energy Diagnostic

A common diagnostic in plasma simulation is to calculate the kinetic energies at the midpoint of the integration interval $t = \frac{\Delta t}{2}$. This can be calculated as follows:

$$W_K = (\gamma - 1)mc^2 = \frac{(\gamma^2 - 1)mc^2}{\gamma + 1} \qquad (36.)$$

where $\gamma$ is evaluated at $t = \frac{\Delta t}{2}$. Substituting equations (19-20) into equation (10) results in:

$$\frac{d\gamma^2}{dt} = \frac{2\widetilde{E}_\parallel}{c^2}\left\{u_\parallel(0) + \widetilde{E}_\parallel t\right\} + \frac{2\widetilde{E}_\perp}{c^2}\left\{u_L(0)\cos\left(\frac{\Omega t}{\gamma_a}\right) + \frac{\gamma_a}{\Omega}\frac{du_L(0)}{dt}\sin\left(\frac{\Omega t}{\gamma_a}\right)\right\} \qquad (37.)$$

where $\frac{du_L(0)}{dt} = \widetilde{E}_\perp - \frac{u_D(0)\Omega}{\gamma_a}$. Integrating over t, one obtains the desired expression:

$$\gamma^2\left(\frac{\Delta t}{2}\right) = \gamma^2(0) + \frac{2\widetilde{E}_\parallel}{c^2}\left\{u_\parallel(0)\frac{\Delta t}{2} + \frac{1}{2}\widetilde{E}_\parallel\left(\frac{\Delta t}{2}\right)^2\right\}$$

$$+ \frac{2\gamma_a\widetilde{E}_\perp}{\Omega c^2}\left\{u_L(0)\sin\left(\frac{\Omega\Delta t}{2\gamma_a}\right) - \frac{\gamma_a}{\Omega}\frac{du_L(0)}{dt}\left[\cos\left(\frac{\Omega\Delta t}{2\gamma_a}\right) - 1\right]\right\} \qquad (38.)$$

Substituting this result into equation (36) gives the desired time-centered energy. This energy value is exact. Most of the time such precision is not needed in a diagnostic. Expanding the trigonometric functions above in a Taylor series gives equation (11), which is most likely sufficient.

3.5 Position equation

Finally, from the expression: $\frac{d\mathbf{x}}{dt} = \frac{\mathbf{u}}{\gamma_a}$, one can integrate to obtain the positions:

$$x_\parallel(t) = x_\parallel(0) + \frac{u_\parallel(0)}{\gamma_a}t + \frac{q}{2m\gamma_a}E_\parallel t^2 \qquad (39.)$$

$$x_L(t) = x_L(0) + \frac{1}{\Omega}u_L(0)\sin\left(\frac{\Omega t}{\gamma_a}\right) - \frac{\gamma_a}{\Omega^2}\frac{du_L(0)}{dt}\left[\cos\left(\frac{\Omega t}{\gamma_a}\right) - 1\right] \qquad (40.)$$

$$x_D(t) = x_D(0) - \frac{\gamma_a}{\Omega^2}\frac{du_L(0)}{dt}\sin\left(\frac{\Omega t}{\gamma_a}\right) - \frac{1}{\Omega}u_L(0)\left[\cos\left(\frac{\Omega t}{\gamma_a}\right) - 1\right] + v_E t \qquad (41.)$$

These equations require that positions and momenta are known at the same time steps. Most PIC codes, however, use a leap-frog scheme where positions and momenta are known at staggered time steps, and it is difficult to make use of these exact equations for position. Therefore, as in [8], we retain the leapfrog scheme for positions:

$$\mathbf{x}(t + \Delta t) = \mathbf{x}(t) + \frac{\mathbf{u}\left(t + \frac{\Delta t}{2}\right)}{\sqrt{1 + \mathbf{u}\cdot\mathbf{u}/c^2},}\Delta t \qquad (42.)$$

For very relativistic plasmas, when most particles are moving near the speed of light, positions errors are small.

## 4.0 Comparison with Exact Analytic Solution

### 4.1 Comparison of Equations

The equations of motion, equation (1), can be expressed in terms of proper time $\tau$ as follows:

$$\frac{d\mathbf{u}}{d\tau} = \gamma\tilde{\mathbf{E}} + \mathbf{u} \times \boldsymbol{\Omega} \text{ and } \frac{d\mathbf{x}}{d\tau} = \mathbf{u} \tag{43.}$$

The exact solution to these equations for fixed **E** and **B** fields, where $\gamma$ was allowed to vary during a time step, was given in covariant form in equations (2.11-2.12) in [8]. The solution is complicated in general, but is relatively simple when $\mathbf{E}_{\|} = 0$:

$$u_{\|}(\tau) = u_{\|}(0) \tag{44.}$$

$$u_L(\tau) = u_L(0)\cos(\tilde{\Omega}\tau) + \frac{1}{\tilde{\Omega}}\frac{du_L(0)}{d\tau}\sin(\tilde{\Omega}\tau) \tag{45.}$$

$$u_D(\tau) = \frac{\tilde{\Omega}}{\Omega}u_L(0)\sin(\tilde{\Omega}\tau) - \frac{1}{\Omega}\frac{du_L(0)}{d\tau}\cos(\tilde{\Omega}\tau) + \gamma(\tau)v_E \tag{46.}$$

where

$$\gamma(\tau) = \gamma(0) + \frac{\tilde{E}_{\perp}}{\tilde{\Omega}c^2}\left\{u_L(0)\sin(\tilde{\Omega}\tau) - \frac{1}{\tilde{\Omega}}\frac{du_L(0)}{d\tau}\left[\cos(\tilde{\Omega}\tau) - 1\right]\right\} \tag{47.}$$

$$\frac{1}{\Omega}\frac{du_L(0)}{d\tau} = \gamma(0)v_E - u_D(0) \text{ and } \tilde{\Omega} = \sqrt{\Omega^2 - \tilde{E}_{\perp}^2 / c^2} \text{ if } \Omega^2 > \tilde{E}_{\perp}^2 / c^2 \tag{48.}$$

One can show by substitution that equations (45-46) satisfy the equations of motion:

$$\frac{du_L}{d\tau} = \gamma\tilde{E}_{\perp} - u_D\Omega \text{ and } \frac{du_D}{d\tau} = u_L\Omega \tag{49.}$$

Comparing equations (20-21) with equations (45-46), one can see that the structure of the solution is the same, but the oscillation frequency is modified and that $\gamma_a$ has been replaced by either $\gamma(0)$ or $\gamma(\tau)$. Equation (24) for $\gamma_a$ is still valid and necessary to convert from time to proper time, but its value is different since $\gamma(\tau)$ is different.

4.2 Comparison of Results for First Benchmark

A version of the code in [8] was written using 3-vector plus time representation. This was done to help understand the physics better and also to provide additional opportunities for optimization. This version, called EAR (Exact Analytic Relativistic), was benchmarked against the version in [8] and produced nearly identical results. Three different versions of the analytic Boris pusher were implemented, using the split-time formulation. They differed only in how accurately the calculation of $\gamma_a$ was performed. The most accurate Boris pusher, called AR (Analytic Relativistic), made use of equation (28), with only one Newton iteration. A version using equation (31) with a fourth order Taylor series, is called A4R, and a version using a second order Taylor series is called A2R. A corrected Boris pusher using equation (14) is called BORISC, and the classic Boris pusher is called BORIS.

For our first benchmark, we selected a test case that has a strong external magnetic field and a moderate electric field with no plasma with the following dimensionless parameters in cartesian coordinates:

$$\tilde{\mathbf{E}}_\perp = [1, -5.0, -1.66] \Longrightarrow \tilde{\mathrm{E}}_\perp = 5.36, \boldsymbol{\Omega} = [0, -25, 75] \Longrightarrow \Omega = 79.1 \qquad (50.)$$

$$\mathbf{u}(\mathbf{0}) = [0, 20, 0], \mathrm{c} = 5 \Longrightarrow \gamma(0) \approx 4 \qquad (51.)$$

In the first case, we compare the two exact pushers, EAR and AR. Normally, we use a time step $\frac{\Omega\Delta t}{\gamma_a} < \pi$ to avoid time aliasing. Here will use $\frac{\Omega\Delta t}{\gamma_a} = 2$, which gives about 3 timesteps per oscillation. Figure 1 shows the time history of $u_D(\tau)$ for EAR and AR, superimposed on a plot for EAR with $\frac{\Omega\Delta t}{\gamma_a} = 0.2$. One can see that EAR and AR are nearly identical. One can also see that EAR gives the same result for both time steps. The maximum value of $\Delta\gamma_a/\gamma_a$ during a step was 2%, so that the assumption that $\gamma$ did not vary during a time step was reasonable.

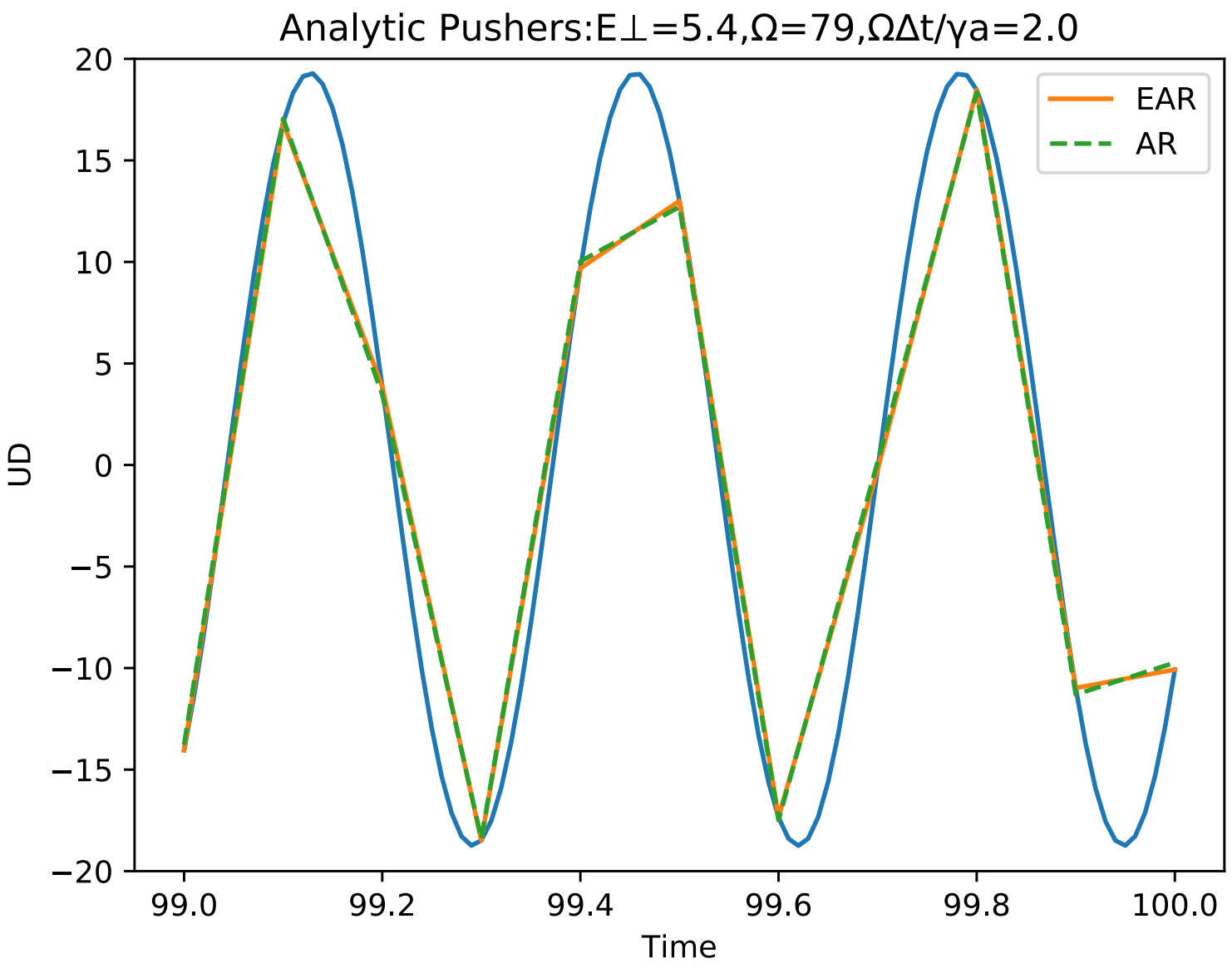

Figure 1: Time history of $u_D(\tau)$ for EAR and AR with $\frac{\Omega\Delta t}{\gamma_a} = 2$ at end of simulation. Solid blue curve shows EAR with $\frac{\Omega\Delta t}{\gamma_a} = 0.2$.

In the second case, we compare EAR and A4R with $\frac{\Omega \Delta t}{\gamma_a} = 0.8$. One can see in Figure 2 that EAR and A4R are nearly identical. Using a larger time step for A4R would show substantial disagreement. The disagreement appears as errors in the phase of the oscillation and in a secular growth of the average value of $\gamma_a$.

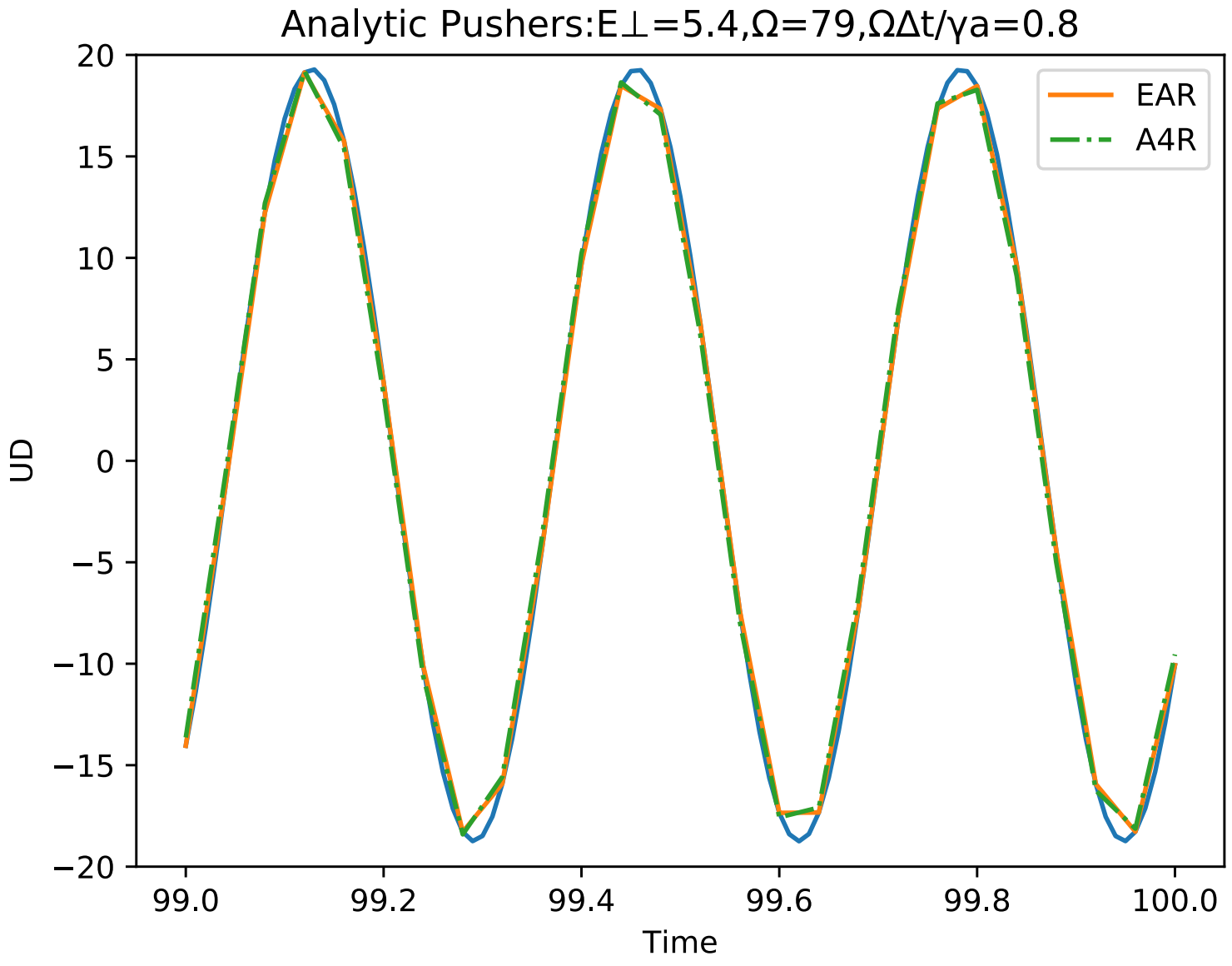

Figure 2: Time history of $u_D(\tau)$ for EAR and A4R with $\frac{\Omega \Delta t}{\gamma_a} = 0.8$ at end of simulation. Solid blue curve shows EAR with $\frac{\Omega \Delta t}{\gamma_a} = 0.2$.

In the third case, we compare EAR and A2R with $\frac{\Omega \Delta t}{\gamma_a} = 0.2$. One can see in Figure 3 that EAR and A2R are nearly identical. Using a larger time step for A2R would show substantial disagreement. In order for the second order BORISC and BORIS versions to obtain the same high precision that the analytic schemes get, one would have to use $\frac{\Omega \Delta t}{\gamma_a} = 0.02$, an order of magnitude smaller time step than normal usage.

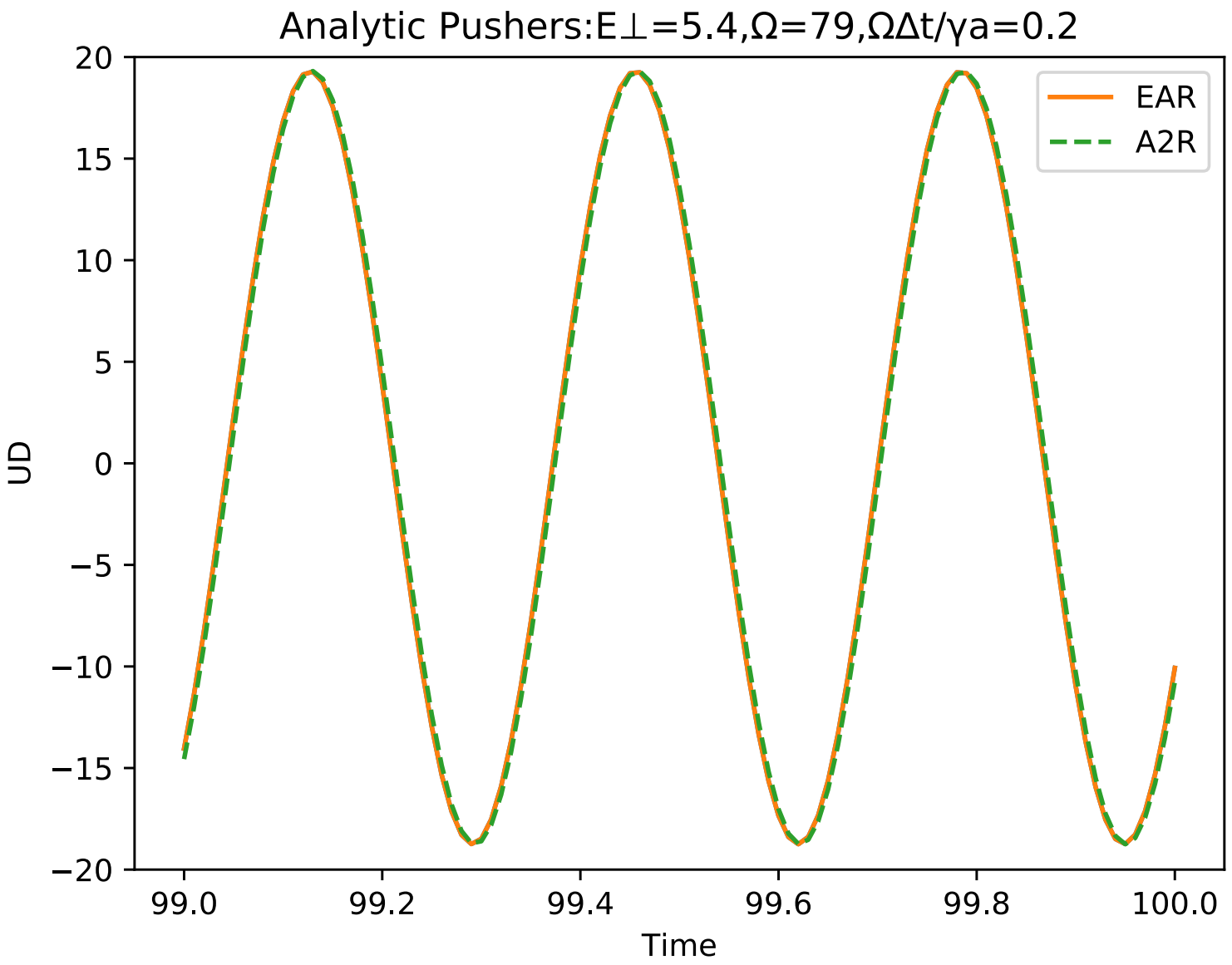

Figure 3: Time history of $u_D(\tau)$ for EAR and A2R with $\frac{\Omega\Delta t}{\gamma_a} = 0.2$ at end of simulation.

### 4.3 Timing Results

To evaluate the computational cost of the various pushers, we measured the CPU time per particle per time step. The particles are 2-1/2D, with 2 position components and 3 velocity components. The measurements were done in double precision with a single core of the 2.3 GHz Intel i9 processor using the Intel Fortran compiler options: ifort -O3 -r8.

| Push Version | CPU Time (nsec) | Time step restriction |
|---|---|---|
| EAR | 174.7 | $\Omega\Delta t / \gamma_a \lesssim \pi$ |
| AR | 100.4 | $\Omega\Delta t / \gamma_a \lesssim \pi$ |
| A4R | 81.2 | $\Omega\Delta t / \gamma_a \lesssim 0.8$ |
| A2R | 67.1 | $\Omega\Delta t / \gamma_a \lesssim 0.2$ |
| BORISC | 46.7 | $\Omega\Delta t / \gamma_a \lesssim 0.02$ |
| BORIS | 35.3 | $\Omega\Delta t / \gamma_a \lesssim 0.02$ |

Table I. CPU Time for various pushers in nsec/particle/time step.

One can see from Table I that the most accurate pusher EAR takes about 5 times more CPU time than the classic Boris pusher. To decide which pusher to use, one must first find the maximum time step possible. This is determined by a maximum frequency $\omega_{max}$ that needs to be resolved in the simulation such as a plasma frequency, $\omega_{max}\Delta t \lesssim 0.2$ or by a Courant condition set by a Maxwell solver, such as $c\Delta t \lesssim 1$. For example, if $\omega_{max} = 1 \Longrightarrow \Delta t \lesssim 0.2$. If $\gamma_a = 4$ and $\Omega = 1 \Longrightarrow \Omega\Delta t/\gamma_a = 0.2$, then one can use the A2R version. However, if $\Omega > 16$, then one should use AR. These time constraints are specifying when the assumption that the electric and magnetic fields are constant during a time step are valid. In all these examples $\Delta\gamma_a/\gamma_a \lesssim 2\%$, so it was always better to use AR rather than EAR.

4.4 Comparison of Results for Second Benchmark

For our second benchmark, we increased the electric field by a factor of 45, the other parameters remained the same:

$$\tilde{\mathbf{E}}_{\perp} = [45, -225, -75] \Longrightarrow \tilde{E}_{\perp} = 241.4, \mathbf{\Omega} = [0, -25, 75] \Longrightarrow \Omega = 79.1 \qquad (52.)$$

Figure 4 shows the time history of $u_D(\tau)$ for EAR and AR with $\frac{\Omega\Delta t}{\gamma_a} = 0.2$, superimposed on a plot for EAR with $\frac{\Omega\Delta t}{\gamma_a} = 2$. One can see that EAR and AR are nearly identical, and that EAR gives the same result for both time steps. However, AR does not give correct results (not shown) for the larger time step because the maximum value of $\Delta\gamma_a/\gamma_a$ during a step was 130%. With the smaller time step, $\Delta\gamma_a/\gamma_a$ was 11%. In this example, it was always better to use EAR than AR because EAR can use a much larger time step. However, if time constraints as described in section 4.3 require use of a smaller time step, then it might be possible that AR is better even in this case.

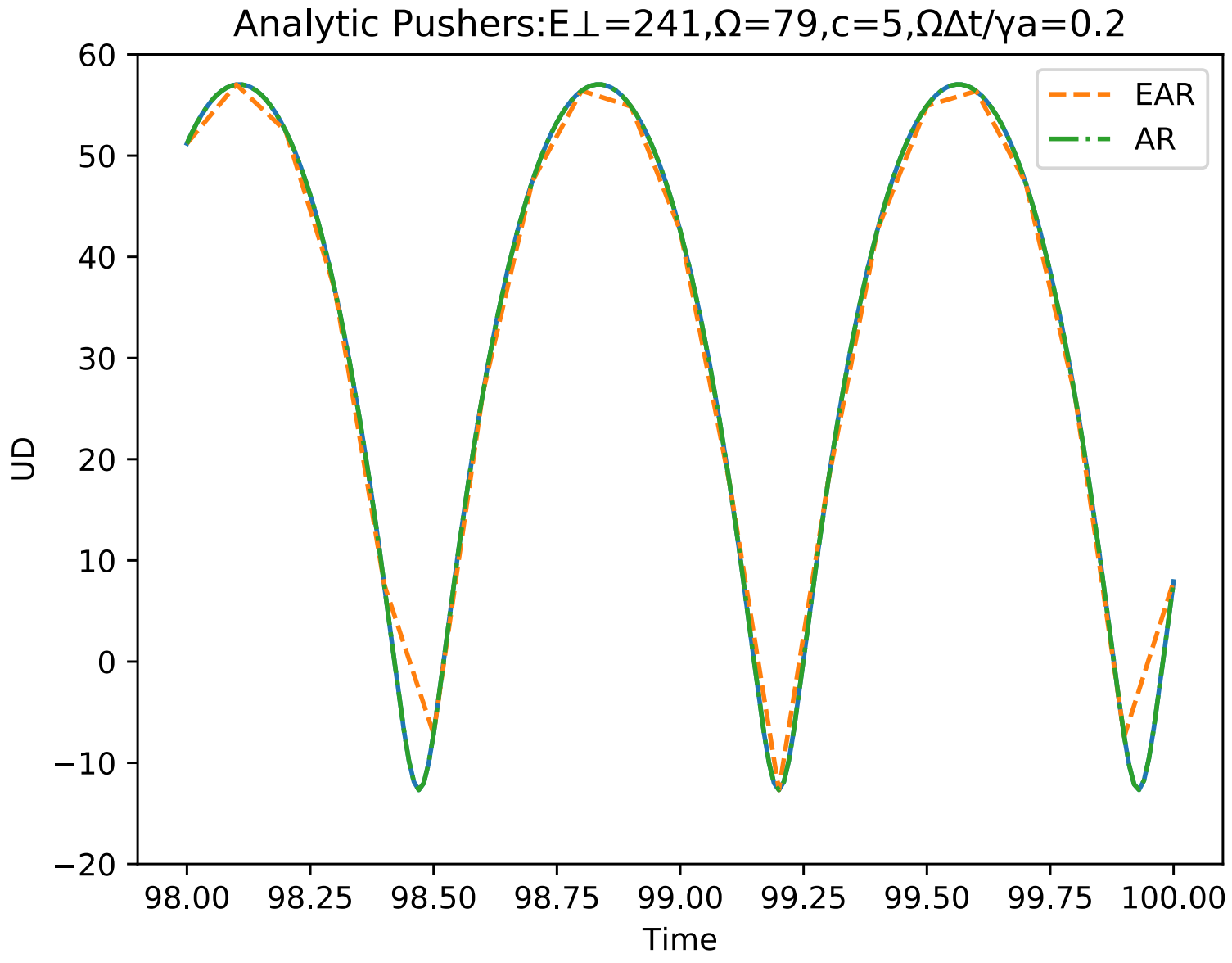

Figure 4: Time history of $u_D(\tau)$ for AR with $\frac{\Omega\Delta t}{\gamma_a} = 0.2$ at end of simulation. Solid blue curve shows EAR with $\frac{\Omega\Delta t}{\gamma_a} = 0.2$ and orange dashed line is EAR with $\frac{\Omega\Delta t}{\gamma_a} = 2.0$

## 5.0 Special Implementation Details

The classic Boris pusher works correctly in the limit of $\Omega \to 0$. Care must be taken with the analytic Boris pushers (AR, A2R, and A4R), however, because they treat $\mathbf{E}_{\parallel}$ and $\mathbf{E}_{\perp}$ differently. If $\Omega = 0$, we set $\hat{\Omega} = 0$, which results in $\mathbf{E}_{\parallel} = \mathbf{0}$ and $\mathbf{E}_{\perp} = \mathbf{E}$, and we reverse equation (35)

$$\frac{\gamma_a}{\Omega}\tan\left(\frac{\Omega\Delta t}{2\gamma_a}\right) \to \frac{\Delta t}{2} \qquad (53.)$$

This leads to the correct result: $\mathbf{u} = \mathbf{u}(\mathbf{0}) + \frac{q}{m}\mathbf{E}\Delta t$.

In addition, in the case of AR, we also need to take the limit of $\Omega \to 0$ when calculating $\gamma_a$. We do this by rewriting equations (27-28) as follows:

$$\gamma(\Delta\tau) = \gamma(0) + \frac{\widetilde{E}_{\parallel}}{c^2}\left\{u_{\parallel}(0)\Delta\tau + \frac{\gamma_a}{2}\widetilde{E}_{\parallel}(\Delta\tau)^2\right\}$$
$$+\frac{\widetilde{E}_{\perp}}{c^2}\left\{u_L(0)\mathrm{sinc}(\Omega\Delta\tau)\Delta\tau - \frac{du_L(0)}{d\tau}\mathrm{cosd}(\Omega\Delta\tau)(\Delta\tau)^2\right\} \quad (54.)$$

$$\Delta t(\Delta\tau) = \gamma(0)\Delta\tau + \frac{\widetilde{E}_{\parallel}}{2c^2}\left\{u_{\parallel}(0)(\Delta\tau)^2 + \frac{\gamma_a}{3}\widetilde{E}_{\parallel}(\Delta\tau)^3\right\}$$
$$-\frac{\widetilde{E}_{\perp}}{c^2}\left\{u_L(0)\mathrm{cosd}(\Omega\Delta\tau)(\Delta\tau)^2 + \frac{du_L(0)}{d\tau}\mathrm{sinc}(\Omega\Delta\tau)(\Delta\tau)^3\right\} \quad (55.)$$

where we define: $\mathrm{sinc}(x) = \frac{\sin(x)}{x}$, $\mathrm{cosd}(x) = \frac{\cos(x)-1}{x^2}$, and $\mathrm{sind}(x) = \frac{\sin(x)-x}{x^3}$. For small arguments we expand them in a Taylor series, which remain finite when $\Omega \to 0$: $\mathrm{sinc}(0) = 1$, $\mathrm{cosd}(0) = \frac{1}{2}$, and $\mathrm{sind}(0) = \frac{1}{6}$. This is not needed for A2R and A4R, since they already rely on Taylor series and work correctly without modification.

We note that for all these analytic pushers, the loop over particles can be easily parallelized with OpenMP directives, since the particles are all independent of each other.

## 6.0 Conclusion

We have derived an analytic version of the classic Boris pusher where the electric and magnetic fields and the particle γ are held fixed during a time step. This pusher occupies a middle ground between the classic Boris pusher and the exact analytic pusher of Fei et. al. in [8], which does not assume that γ is held fixed during a time step. We have shown when such a pusher is advantageous and when it is not.

## Acknowledgments

This work was supported by USDOE SciDAC grant DE-AC02-07CH11359. We wish to thank Jean-Noel Leboeuf and Richard Sydora for their helpful comments in their critique of this work.